# Solitons: from Charge Density Waves to FFLO in superconductors.

*S. Brazovskii*
*CNRS, UMR 8626 LPTMS, Univ. Paris-Sud, Orsay, France*

This short review aims to summarize on "What the Charge Density Waves can tell to other inhomogeneous states in strongly correlated systems, particularly to spin-polarized superconductors". We shall update on expanding observations of solitons in quasi 1D CDW conductors and link them to the growing information and demands related to inhomogeneous spin-polarized states in superconductors. The related theory, existent or awaited for, stretches from solitons in 1D models to vortex-like elementary excitations in 2D,3D ordered incommensurate CDWs and superconductors.

1. **Introduction.**

Both Charge Density Waves (CDW) and superconductors (SC) are the spin-singlet states, which are subject to deparing under the Zeeman splitting effect of the magnetic field [1]. Above a certain critical field, the ground state develops a periodic superstructure, see a short review and references in [2,3]. In superconductors it is known as the FFLO state, which has attracted a great deal of attention recently - see e.g. [4,5], because of events in organic and heavy-fermion superconductors, in cold atoms. Its formation is expected to be a very week effect in conventional superconductors, unless it is endorsed by strong coupling $\Delta \sim E_F$, or by open or partly flattened Fermi surfaces. Then the theories of early 80's (S.B. et al for CDWs, Buzdin et al, Machida et al for SCs) - see [2], predict formation of solitonic lattices with unpaired spins localized at midgap states near the order parameter nodes. For CDWs there are convincing theoretical and experimental evidences [3] that beyond the coherent walls, also the separate amplitude solitons (the walls building blocks) exist as quasi particles - the spinons. These subgap particles are more favourable than electrons, and they would determine the observable properties which are usually ascribed to conventional electronic over-gap excitations. At presence of 2D or 3D long range coherence, these topologically nontrivial solitons experience the confinement resulting in the spin-charge recombination. It originates the symmetry broken roton configuration for the phase coupled with the spin-bearing kink in the core. Based upon the CDW notion, in SCs we expect to find a tightly bound pair of half-integer vortices sharing one unpaired spin. This state seems to win over its close alternative in 2D: a single pancake vortex with one half-filled intra-gap state.

2. **Observations of solitons.**

We start to show most typical and convincing evidences for existence of the gap in the excitation spectrum of singlet-ground-state electronic systems: superconductors and CDWs, Fig.1. Recall the standard BCS - Bogolubov view on the nature of what is seen by tunneling and by other quasi-particles related experiments. States are the linear combinations of: electrons and holes at $\pm\mathbf{p}$ for SC, or of electrons at $-\mathbf{p}$ and $\mathbf{p}+2\mathbf{p}_f$ for CDWs. Corresponding spectra are $E(\mathbf{k}) = \pm(\Delta^2 + (v_f k)^2)^{1/2}$, $\mathbf{k} = \mathbf{p} - \mathbf{p}_f$.
<u>But is it always true?</u> It is proved to be "yes" for typical SCs; but questionable for strong coupling cases: High-$T_c$, real space pairs, cold atoms, bi-polarons; clearly incomplete for CDWs as proved by modern experiments; certainly inconsistent for 1D and even quasi 1D systems as proved theoretically, see [3]. These are the solitons and their arrays which are responsible for these confusions. Let us summarize a complex of well established facts for the incommensurate CDWs (ICDW) which are symmetrically equivalent to SCs.

The CDW, at first sight, is a semiconductor with free electrons or holes near the gap edges $\pm\Delta_0$. Then the same gap is expected to perform the following functions: **1.** $\Delta_0$ - in kinetics and thermodynamics (conductivity, spin susceptibility, heat capacitance, NMR); **2.** $\Delta_0$ - in dynamics (photoemission, external tunneling; **3.** $2\Delta_0$ -in optics or in internal tunneling; **4.** $\Delta_0$ – as a threshold for electronic pockets formation by doping or injection (FET).

But nothing of this standard picture takes place in ICDWs: **1.** Activation energies from transport in directions on-chain and inter-chain differ by several times (TaS$_3$ or Blue Bronze : 200K and 800K); **2.** Activations for spins and from relaxation are in-between - 600K; **3.** Optical absorption peaks at $2\Delta_0$, but is deeply spread below; **4.** Thresholds for charge transfer are as low as the on-chain activation, i.e. as the interchain decoupling scale $T_c$; **5.** Charge injection is accommodated into the extended ground state via phase slip processes, rather than via formation of Fermi pockets.

This decade, a true workshop on solitons was opened in organic conductors like (TMTCF)$_2$X, see [7]. The facility is provided by the discovery of the ferroelectricity endorsed by the charge ordering (*Nad, Monceau, and S.B.; S. Brown et al*). The zoo of solitons is largely accessed thanks to possibility of switching on/off of the Mott state by means of the charge ordering. Here we recall only one example: optical evidences for creation of solitons' pairs and their bound states Fig.2.

While the charge ordering in the organic conductor is a crystal of electrons, the conventional CDW is a crystal of electron pairs. Its lowest energy current carrier may be the charge-2e defect of adding/missing one period at the defected chain. It is the $\pm 2\pi$ soliton of the ICDW order parameter $\Psi_{ICDW}= A\cos(2K_f x+\varphi)$, which has been recently captured and visualized in STM experiments [8], Fig.3.

What comes if the singlet pair of the CDW is broken into spin ½ components? Unlike the quotations related to the Fig.1, it will <u>not</u> be an expectedly liberated electron-hole pair at $\pm\Delta_0$, but rather two spin carrying "amplitude solitons" – zeros of the order parameter distributed over the length $\xi_0$. The unpaired electron is trapped at the midgap state of the amplitude soliton, with the energy $\approx 2\Delta_0/3$, the total charge 0, and the spin ½. This is the CDW realization of the spinon, in a similarity to the neutral kink in the polyacetylene. The regular lattice of these solitons might have been observed as the CDW superstructure in high magnetic field (HMF) [1] and in spin-Peierls systems in HMF (as clearly seen by the NMR [9]).

Generalization of the solitonic lattice from the ICDW to the SC is the FFLO phase in spin-polarized superconductors. Then the same AS is an elementary stripe fragment in both cases of the CDW and FFLO, with a similarity [3] to holons in doped AFM insulator.

Can we see the half-period AS as good as we could see the full-period phase soliton, Fig.3? Fortunately, there is an ill-noticed success for a dimerized system of a transition-metal-chalcogenide chain [10]. Commenting to the STM picture reproduced in Fig.4, the authors claimed: «*For the first time the spin soliton has been visualized in real space*». While the solitons cannot be always visualized, they show themselves in spectral features as in optics, Fig.2, or by tunneling as in Fig.5.

3. **Solitons and the long range order.**

The major puzzle, as well as the inspiration, coming from the above quoted tunneling and STM experiments is that the amplitude solitons were observed within the low temperature (T<$T_c$) phase with the long range order. The hidden obstacle is the effect of the confinement appearing in higher dimensions D>1 [3]. Commuting between degenerate minima on one chain would lead to a loss of the interchain ordering energy ~ "length to the next defect". Other modes need to be activated to cure the topological defect.

At the 1D level, the Amplitude Soliton $\Psi$ (x=-$\infty$) =-$\Psi$ (x=$\infty$) performs the amplitude sign change **A → -A** at arbitrary $\varphi$ =**cnst**. It is favorable in energy in comparison with an electron, but prohibited to be created dynamically even in 1D, and prohibited to exist even stationary at D>1. The resolution is to invoke the combined symmetry: the amplitude kink **A → -A** coupled with the half-integer φ → φ+π vortex of the phase rotation which compensates for the amplitude sign change. The resulting Spin-Roton complex allows for several interpretations. 2D view: a pair of π-vortices shares the common core bearing one unpaired spin which stabilizes the state. 3D view: ring of a half-flux vortex line, its center confines the spin. Today's perspective: nucleus of the melted FFLO phase in the spin-polarized SC.

Recall finally an alternative microscopic insight to excitations in these spin-gap cases – SC or CDW. The starting single chain level is well described by the bosonisation language. The Hamiltonian $H_{1D} \sim (\partial\theta)^2 - V\cos(2\theta) + (\partial\varphi)^2$ is written in terms of the spin - θ and the charge - φ phases. The energy **V** comes from the backward exchange scattering $g_1$ of electrons. The pair-breaking excitation - the **s=1/2** spinon, is the soliton connecting the degenerate minima of $H_{1D}$ : **θ → θ+π**. The singlet order parameter, for either SC or CDW (depending on a definition of the charge phase φ) is like **Ψ**$_{SC,CDW}$**~ cosθ exp(i φ)**. Its amplitude cosθ changes the sign across the allowed π soliton, hence the spinon is an alternative description of the same amplitude soliton which appears in BCS-Peierls type models.

4. **FFLO phase in superconductors.**

FFLO refers to an undulating phase in superconductors with an imbalanced spin population: In this abbreviation, FF and LO stand for Fulde&Ferrell 1964 and Larkin&Ovchinnikov 1964 articles. These authors challenged the standard picture of a superconductor with a small spin polarization. The conventional homogeneous phase implies filling the excess spins to quasi-particle states above the gap, according to the Fig.6a copied from the original FF publication.

Modulated phases of the complex order parameter Ψ with a wave number **Q≠0** have been suggested: FF: **Ψ~exp(iQx)**, LO: **Ψ~cos(Qx)**. The appropriately chosen vector Q (Fig.6b) erases mismatching at some (at all in the quasi-1D case) parts of the FS, hence preventing the collapse of the SC phase in the magnetic field. While this interpretation is valid for both suggestions FF and LO, there is a particular insight to the LO case relevant to solitonic lattices. The planes of the order parameter zeros are able to concentrate the excess spins providing the split intragap states which are able to accommodate unpaired electrons. This scenario is directly linked to the solitonic lattices in quasi-1D case. The available exact solution for the FFLO phase in quasi 1D system is shown at the Fig.7.

5. **Inverse rout: from stripes to solitons and fractional vortices.**

If the solitonic lattice melts, then in 1D each element becomes a particle – the amplitude soliton = spinon. In D=2, the amplitude defect should be complemented by the pair of π-vortices, Fig.8. This quasi 1D version is a secure generalization of the rigorous 1D picture. But for how far can we extrapolate to general superconductors which do not possess a strong anisotropy? In general, the cost of creating a pair of vortices is ~$E_{ph}$**Log(L)**, where $E_{ph}$ is a characteristic energy of phase deformations and the string length L is the distance (in lattice units) between the opposite π-vortices. This loss must be equilibrated by the gain -Δ'L for the string formation, where Δ'~Δ is the energy gained from accommodating unpaired electrons to the midgap states. In the quasi 1D case, $E_{ph}$~$T_c$<Δ is given by the low phase ordering temperature $T_c$, then the total energy $E_{ph}$**Log(L) -Δ'L/a** keeps to be negative down to smallest atomic length **L~a** – this is why the combined kink-roton complex is certainly a stable quasi-particle. But for isotropic SCs, $E_{ph}$~$E_F$ which allows for only a large scale complex, at **L/a> $E_F$/Δ'**. The strong coupling limit Δ~$E_F$ is necessary, which leaves this scenario for a bipolaronic SC or for a condensate of paired cold Fermi atoms.

In absence of a microscopic theory for strong coupling vortices (i.e. with only a single pair of intra-gap states), we can rely upon existing [12] numeric modeling (still done within the BCS scheme). And the results are supporting indeed: at presence of unpaired spins, the usual integer $2\pi$ vortex, created by rotation (magnetic field), splits into two $\pi$- vortices, Fig. 9.

The energetics behind the vortex splitting is understandable: for any $2\pi N$ vortex, the energy $\sim N^2$ hence ½ of it can be gained by splitting in 2 vortices with vorticities $\pi N$ – this is why there is no such a thing as $4\pi$, etc. vortex. But splitting of **N=1** vortex into two ½ ones is commonly prohibited because there is no self-mapping at noninteger **N**. It becomes allowed if the amplitude domain wall opens between the split cores. The node in order parameter amplitude allows for the "prohibited" **N=1/2** circulation, hence forming a pair of two ½ vortices connected by the spin-carrying amplitude domain wall .

At first sight, there is a simper construction – a competitor to the scenario of the kink-roton complex. This is a single integer vortex with a half-filled intra-gap core level – an extension of the Caroli-DeGennes-Matricon staircase to the smallest (two) number of levels. But it does not seem to work: unlike the zero-level midgap states originated by the amplitude kink, the vortex core levels are split and repelled (with increasing coupling) towards the gap edges; so the energy gain from the localization of one unpaired electron is expected to be small as $\Delta' \sim \Delta_0^2/E_F$.

## 6. Conclusions.

Existence of solitons is proved experimentally in single- or bi-electronic processes of CDWs in several quasi 1D materials. They feature self-trapping of electrons into mid-gap states and separation of spin and charge into spinons and holons, sometimes with their reconfinement at essentially different scales. Continuously broken symmetries allow for solitons entering the D>1 world of long range ordered states: SC, ICDW, SDW. Solitons take forms of amplitude kinks which are topologically bound to semi-vortices of gapless modes – half integer rotons. These combined particles substitute for electrons - certainly in quasi-1D systems, which is valid for both charge-and spin-gaped cases. The description is extrapolatable to strongly correlated isotropic cases. Here it meets the picture of fragmented FFLO or stripe phases.

**Acknowledgements.**

A support was provided by the ANR program (the project BLAN07-3-192276).

This article summarizes the presentations given at the conferences STRIPES 2008 and ECRYS 2008 [13].

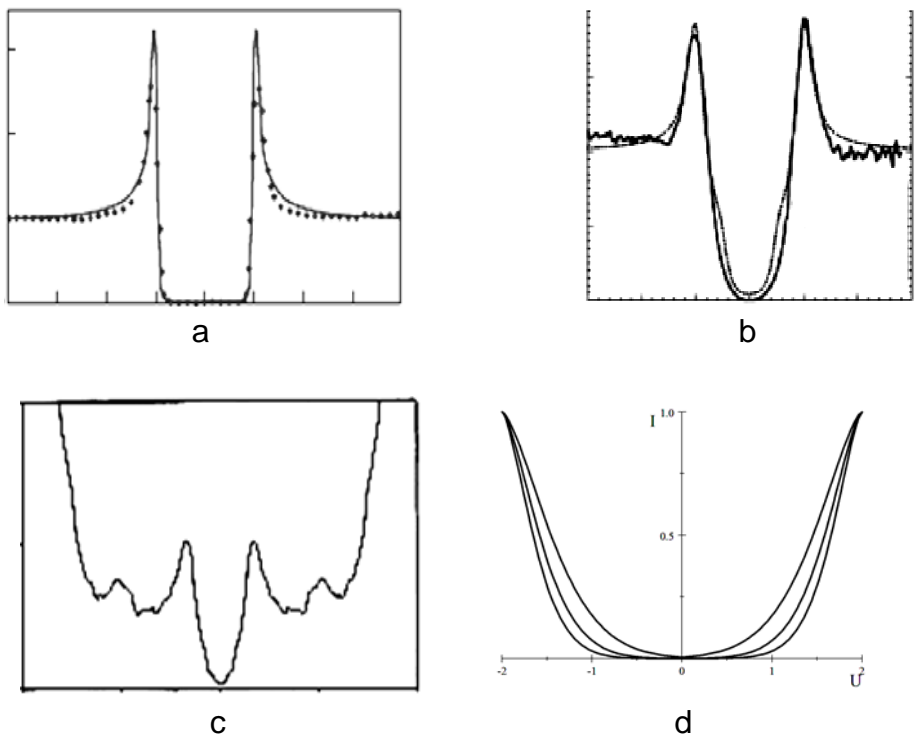

Figure 1. Deparing gaps from tunneling experiments – plots of the tunneling conductance dI/dU versus voltage U. Superconductors: Nb (a) and $CaC_6$ (b); CDW in $NbSe_3$ (c).
(d) - theoretical prediction [6] for the instanton-mediated tunneling current I(U) in the subgap region $|U|<2\Delta$. (U is shown in units of $\Delta$; the plots correspond to temperatures $T/\Delta =1/4, 1/6, 1/8$).

---

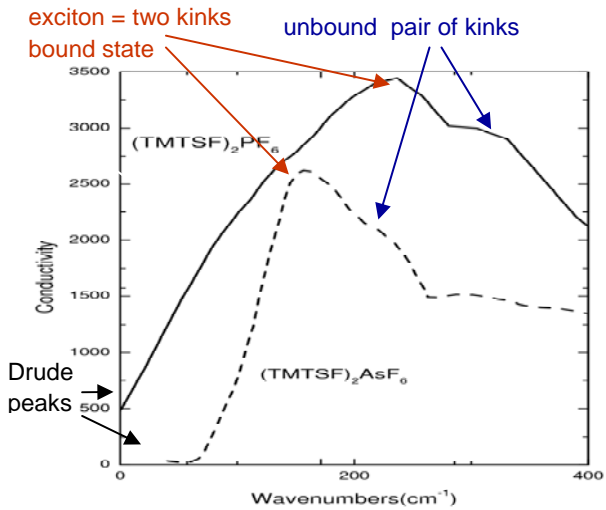

Figure 2. Interpretation of optics of conducting $(TMTSF)_2X$ in terms of expectations for the Charge Ordering (Mott insulator) state. (Using results by M. Dressel and L. Degiorgi groups).

---

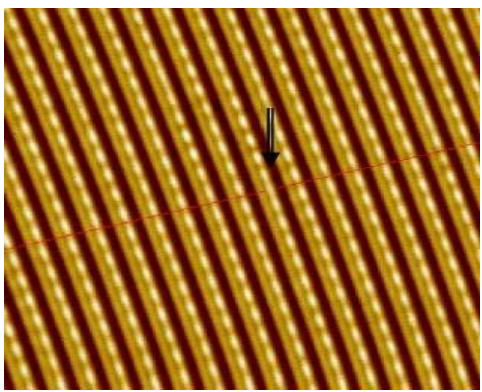

Figure 3. Visualization of the $2\pi$ soliton – the prefabricated pair of electrons, by the STM on $NbSe_3$ [8]. At the (red) front line the defected chain is displaced by half of the period. Along the defected chain the whole period $\pm 2\pi$ is missed or gained – a pair of electrons or holes is accommodated to the ground state.

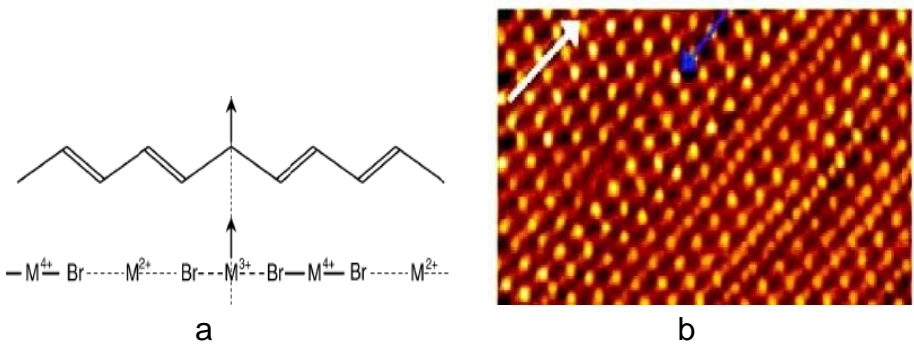

a    b

Figure 4. Quasi1D Halogen-bridged complex $Ni_{0.05}Pd_{0.95}Br$ and the STM visualization of the half-period soliton. White arrow: 1D chains direction, blue arrow: chain with the defect [10].

---

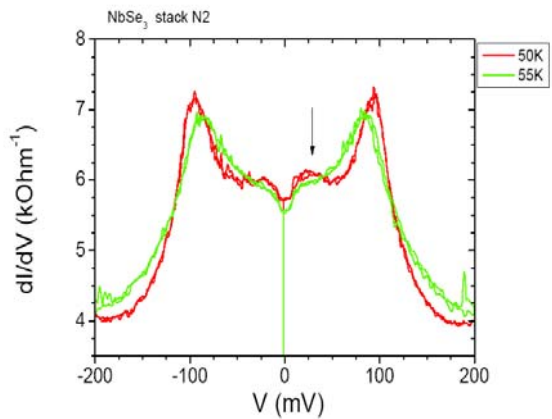

Figure 5. Tunneling in mesa-junctions of $NbSe_3$. The complex of tunneling features shows a coexistence of several spectral processes: peak $2\Delta$ for inter-gap creation of e-h pairs, creation of the amplitude soliton at $E_{as}=2\Delta/3$, bi-particle channel at $V_t \ll \Delta$ – the spinless charge injection threshold [11].

---

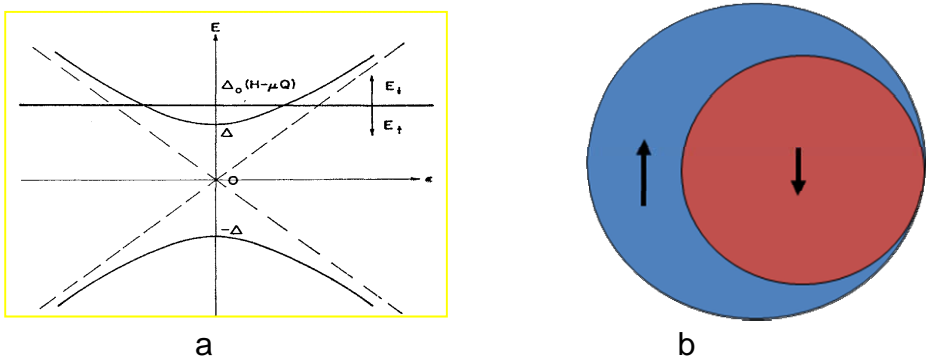

a    b

Figure 6. Spin-imbalanced superconductor. a) Filling of the bare spectrum keeping the homogeneous phase. b) Modulated phase improves the matching at some parts of up/down Fermi surfaces.

---

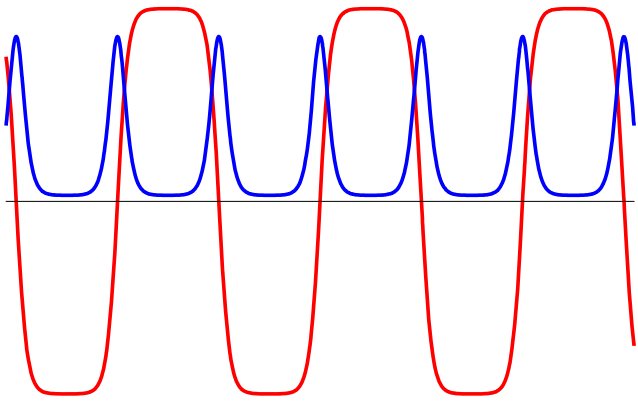

Figure 7. Solitonic lattice in CDW or SC under slightly supercritical Zeeman splitting. The plots show distributions of the order parameter and of the density of unpaired spins - mid-gap states concentrated near the gap zeros.

---

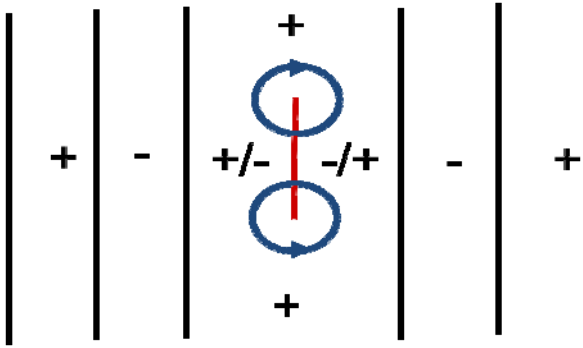

Figure 8. Kink-roton complexes as nucleuses of a melted FFLO lattice. The defect is embedded into the regular stripe structure (black lines). +/- are the alternating signs of the order parameter amplitude. Termination points of a finite segment L (red color) of the Ψ=0 line must be encircled by semi-vortices of the π rotation (blue circles) to resolve the signs mismatch. The minimal segment corresponds to the elementary kink carrying spin 1/2.

---

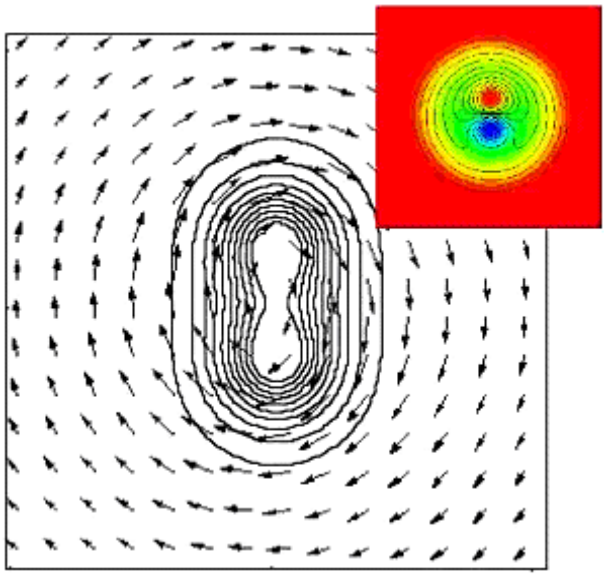

Figure 9. Splitting of the conventional integer orbital vortex into two counterparts in presence of a population of unpaired spins, taken from [12]. For our goals, we just reformulate these results inversely – unpaired spins create the vortex pair even at no orbital magnetic field.